\documentclass[leqno,12pt,twoside]{article}
\usepackage{hyperref}
\usepackage{amsmath}
\usepackage[psamsfonts]{amssymb}
\usepackage{amsthm}
\usepackage[mathscr,psamsfonts]{eucal}
\usepackage{cmmib57}
\DeclareFontFamily{U}{eur}{\skewchar\font'177}
\DeclareFontShape{U}{eur}{m}{n}{
  <-6> eurm5 <6-8> eurm7 <8-> eurm10
  }{}
\DeclareFontShape{U}{eur}{b}{n}{
  <-6> eurb5 <6-8> eurb7 <8-> eurb10
  }{}
\DeclareMathAlphabet{\eurm}{U}{eur}{m}{n}
\DeclareMathAlphabet{\eubf}{U}{eur}{b}{n}
\DeclareFontFamily{OML}{cyr}{}
\DeclareFontShape{OML}{cyr}{m}{n}{
   <5> <6> <7> <8> <9>
   <10> <10.95> <12> <14.4> <17.28> <20.74> <24.88> wncyr10
  }{}
\DeclareSymbolFont{rusletters}{OML}{cyr}{m}{n}
\DeclareSymbolFontAlphabet{\rusmath}{rusletters}
\DeclareMathSymbol{\cyq}{\rusmath}{rusletters}{"71}
\newcounter{assump}
\newtheorem{Assumption}{\indent Assumption}[assump]
\newtheoremstyle{MyThm}
  {3pt}
  {3pt}
  {\itshape}
  {\parindent}
  {\bfseries}
  {.}
  {.5em}
  {}
\swapnumbers
\theoremstyle{MyThm}
\newtheorem{Definition}{Definition}[section]
\newtheorem{Lemma}[Definition]{Lemma}
\newtheorem{Proposition}[Definition]{Proposition}
\newtheorem{Theorem}[Definition]{Theorem}
\newtheorem{Corollary}[Definition]{Corollary}
\newtheorem{Remark}[Definition]{Remark}
\newtheorem{Note}[Definition]{Note}
\newtheorem{Example}[Definition]{Example}
\newcommand{\bAs}{\begin{Assumption}\em}
\newcommand{\eAs}{\end{Assumption}}
\newcommand{\bDf}{\begin{Definition}\em}
\newcommand{\eDf}{\end{Definition}}
\newcommand{\bLm}{\begin{Lemma}}
\newcommand{\eLm}{\end{Lemma}}
\newcommand{\bPr}{\begin{Proposition}}
\newcommand{\ePr}{\end{Proposition}}
\newcommand{\bTh}{\begin{Theorem}}
\newcommand{\eTh}{\end{Theorem}}
\newcommand{\bCr}{\begin{Corollary}}
\newcommand{\eCr}{\end{Corollary}}
\newcommand{\bRm}{\begin{Remark}\em}
\newcommand{\eRm}{\end{Remark}}
\newcommand{\bNt}{\begin{Note}\em}
\newcommand{\eNt}{\end{Note}}
\newcommand{\bEx}{\begin{Example}\em}
\newcommand{\eEx}{\end{Example}}
\newcommand{\bPf}{\begin{proof}[\noindent\indent{\sc Proof}]}
\newcommand{\ePf}{\renewcommand{\qedsymbol}{}\end{proof}}
\newcommand{\bEq}{\begin{eqnarray}}
\newcommand{\eEq}{\end{eqnarray}}
\newcommand{\beq}{\begin{eqnarray*}}
\newcommand{\eeq}{\end{eqnarray*}}

\newcommand{\myskip}{\vspace*{8pt}}
\newcommand{\R}{I\!\!R}

\newcommand{\Cin}{C^\infty}
\newcommand{\h}{\hbar}
\newcommand{\rmi}{\eurm{i}\,}

\newcommand{\mto}{\mapsto}

\newcommand{\der}{\partial}
\newcommand{\nab}{\nabla}

\newcommand{\Upa}{^{\uparrow}}

\newcommand{\Nat}{^{\natural}}
\newcommand{\Fla}{^{\flat}}

\DeclareMathOperator{\con}{\lrcorner}
\newcommand{\com}{\circ}
\newcommand{\car}{\times}
\newcommand{\ten}{\otimes}

\newcommand{\wed}{\wedge}

\newcommand{\ucar}[1]{\underset{#1}{\times}}

\newcommand{\uten}[1]{\underset{#1}{\otimes}}

\newcommand{\sqrtg}{\sqrt{|g|}}
\newcommand{\sqrtgo}{\frac{\der_0 \sqrtg}{\sqrtg}}

\def\QED{\hskip0.1em\hfill\null\
\null\nobreak\hfill\kern3pt\vbox{\hrule\hbox
   {\vrule\kern1pt\vbox{\kern1.7pt\hbox{$\scriptscriptstyle{QED}$}
    \kern0.2pt}\kern1pt\vrule}\hrule}}

\def\END{\hskip0.1em\hfill\null\
\null\nobreak\hfill\kern3pt\vbox{\hrule\hbox
  {\vrule\kern1pt\vbox{\kern1.7pt\hbox{$\,\,\,\vspace{5pt}$}
    \kern0.2pt}\kern1pt\vrule}\hrule}}
\DeclareMathOperator{\byd}{\,{\raisebox{.1ex}{$\eurm :$}{\eurm =}}\,}

\DeclareMathOperator{\id}{id}
\DeclareMathOperator{\Div}{div}

\newcommand{\bE}{\boldsymbol{E}}

\newcommand{\bH}{\boldsymbol{H}}

\newcommand{\bQ}{\boldsymbol{Q}}

\newcommand{\bS}{\boldsymbol{S}}
\newcommand{\bT}{\boldsymbol{T}}

\newcommand{\EL}{\eurm{L}}

\newcommand{\ES}{\eurm{S}}

\newcommand{\cA}{\mathcal{A}}

\newcommand{\cE}{\mathcal{E}}

\newcommand{\cH}{\mathcal{H}}

\newcommand{\cL}{\mathcal{L}}

\newcommand{\cO}{\mathcal{O}}
\newcommand{\cP}{\mathcal{P}}

\newcommand{\BC}{{\mathbb{C}}}

\newcommand{\BL}{{\mathbb{L}}}
\newcommand{\BM}{{\mathbb{M}}}

\newcommand{\BT}{{\mathbb{T}}}

\newcommand{\alp}{\alpha}

\newcommand{\gam}{\gamma}
\newcommand{\del}{\delta}

\newcommand{\lam}{\lambda}

\newcommand{\ome}{\omega}

\newcommand{\The}{\Theta}
\newcommand{\Lam}{\Lambda}

\newcommand{\Ome}{\Omega}
\newcommand{\obf}{\overset{o}{f}}
\newcommand{\obnab}{\overset{o}{\nabla}}
\newcommand{\obDel}{\overset{o}{\Delta}}
\newcommand{\banab}{\bar{\nabla}}
\newcommand{\baG}{\bar{G}}
\newcommand{\baP}{\bar{P}}
\newcommand{\venab}{\nab\!\!\!\!^{^{^\vee}}}

\newcommand{\lsqu}{[\![}
\newcommand{\rsqu}{]\!]}
\title{
{\bf Comparison between
\\
Geometric Quantisation and
\\
Covariant Quantum Mechanics}
}
\author{
{\bf Marco Modugno}
\smallskip
\\
{\small University of Florence}
\\
{\small Department of Applied Mathematics}
\\
{\small Via S. Marta 3, 50139 Florence, Italy}
\\
{\small email: {\tt modugno@dma.unifi.it}}
\bigskip
\\
{\bf Carlos Tejero Prieto}
\smallskip
\\
{\small University of Salamanca}
\\
{\small Department of Mathematics}
\\
{\small Plaza de la Merced 4, Salamanca, Spain}
\\
{\small email: {\tt carlost@gugu.usal.es}}
\bigskip
\\
{\bf Raffaele Vitolo}
\smallskip
\\
{\small University of Lecce}
\\
{\small Department of Mathematics `E. De Giorgi'}
\\
{\small Via per Arnesano, 73100 Lecce, Italy}
\\
{\small email: {\tt Raffaele.Vitolo@unile.it}}
}
\begin{document}
\maketitle
\begin{abstract}

We compare the covariant formulation of Quantum Mechanics on a curved
spacetime fibred on absolute time with the standard Geometric
Quantisation.
\end{abstract}
\vfill


\newpage
\tableofcontents
\newpage
\section{Introduction}

Some years ago A. Jadczyk and M. M. proposed a covariant formulation
of Quantum Mechanics for a scalar particle on a curved spacetime with
absolute time, based on non standard methods such as fibred manifolds,
jet spaces, non--linear connections, systems of connections,
cosymplectic structures and Froelicher smooth spaces
\cite{JadMod92,JadMod94a,JadMod94b}. This theory has been extended to
spin particles in cooperation with D. Canarutto \cite{CanJadMod95},
further developed in cooperation with J. Jany\v{s}ka, D. Saller, C.
Tejero Prieto and R. Vitolo
\cite{JadJanMod98,Jan95a,Jan95b,Jan98,Jan00,JanMod96b,JanMod99,
Mod98,Mod00,ModTejVit99,ModVit96,
SalVit99,Vit96a,Vit96b,Vit99a,Vit99b} and partially extended to a
Lorentz manifold in cooperation with J. Jany\v{s}ka and R. Vitolo
\cite{JanMod96a,JanMod96b,JanMod97a,JanMod97b,Vit00}.

In the proceedings of the previous session of the meeting on Lie
Theory, we have accounted for a summary of this theory \cite{Mod98}.
In order to capture the non standard methods and results of this
theory it would be useful to compare it with the more standard
Geometric Quantisation. This is the goal of the present paper.

For the sake of simplicity, our theory in the Galileian case will be
conventionally referred to as {\em Covariant Quantum Mechanics\/}
(CQM). Moreover, we shall be concerned with the main thread of {\em
Geometric Quantisation\/} (GQ) and omit to consider special approaches
dealing with quantisation of cosymplectic structures 
\cite{deLMarPad97} and so on. We have no pretension at all of
analysing extensively the wide literature of Geometric Quantisation;
such a task would require a much larger space than a short note.
Here, we just try to discuss some basic items concerning the
comparison of the above theories.

\myskip

{\bf Acknowledgements}: The first author thanks the organizers of the
Meeting for the invitation and for the warm hospitality. This work
has been partially supported by University of Florence, Italian MURST
and Italian GNFM of CNR.
\section{Covariant Quantum Mechanics}
\label{Covariant Quantum Mechanics}

Let us start with a very brief sketch of the skeleton of the theory.
For further details the reader can refer, for instance, to
\cite{JadMod94b,JadJanMod98,Mod98}.

The covariance of the theory includes also independence from the
choice of units of measurements. For this reason, we need a rigorous
treatment of this feature and assume the following ``positive
1--dimensional semi--vector spaces" over $\R^+$ as fundamental unit
spaces (roughly speaking they have the same algebraic structure of
$\R^+$, but no distinguished generator over $\R^+$): the space
$\BT$ of {\em time intervals\/}, the space $\BL$ of {\em lengths\/},
the space $\BM$ of {\em masses\/}.

Moreover, we assume the {\em Planck constant\/} to be an element
$\h \in \BT ^* \ten \BL^2 \ten \BM$.

We refer to a particle with mass
$m \in \BM$ and charge $q \in \BT^* \ten \BL^{3/2} \ten \BM^{1/2}$.
\subsection{Classical theory}
\label{Classical theory}

The classical framework is described in the following way.

The {\em spacetime\/} is an oriented $(n+1)$--dimensional manifold
$\bE$ (in the standard case $n = 3$), the {\em absolute time\/} is an
affine space associated with the vector space $\R \ten \BT$, the {\em
absolute time map\/} is a fibring $t : \bE \to \bT$. We denote
fibred charts of spacetime by $(x^\lam) \equiv (x^0, x^i)$. The
tangent space and the vertical space of $\bE$ are denoted by $T\bE$
and $V\bE$.

A {\em motion\/} is a section $s : \bT \to \bE$. The
{\em phase space\/} is the first jet space of motions $J_1\bE$
\cite{KolMicSlo93,ManMod83,Sau89}.  We denote fibred charts of
phase space by
$(x^0, x^i; x^i_0)$. The {\em absolute velocity\/} of a motion
$s$ is its first jet prolongation $j_1s : \bT \to J_1\bE$. An
{\em observer\/} is a section $o : \bE \to J_1\bE$ and the
{\em observed velocity\/} of a motion $s$ is the map
$\nab[o] s \byd j_1s - o \com s : \bT \to \BT^* \ten V\bE$.

The {\em spacelike metric\/} is a scaled Riemannian metric of the
fibres of spacetime
$g : \bE \to \BL^2 \ten (V^*\bE \uten{\bE} V^*\bE)$. Given a particle
of mass $m$, it is convenient to consider the re--scaled
spacelike metric
$G \byd \tfrac{m}{\h} \, g : \bE \to \BT \ten (V^*\bE \uten{\bE}
V^*\bE)$.

The {\em gravitational field\/} is a time preserving torsion
free linear connection of the tangent space of spacetime
$K\Nat : T\bE \to T^*\bE \uten{T\bE} TT\bE$, such that
$\nab[K\Nat] g = 0$ and the curvature tensor
$R[K\Nat]$ fulfills the condition
$R\Nat{}_\lam{}^i{}_\mu{}^j = R\Nat{}_\mu{}^j{}_\lam{}^i$.

The {\em electromagnetic field\/} is a scaled 2--form
$f : \bE \to (\BL^{1/2} \ten \BM^{1/2}) \ten \Lam^2 T^*\bE$, such
that $df = 0$. Given a particle of charge $q$, it is convenient to
consider the re--scaled electromagnetic field
$F \byd \tfrac{q}{\h} \, f : \bE \to \Lam^2 T^*\bE$.

The electromagnetic field $F$ can be ``added", in a covariant way, to
the gravitational connection $K\Nat$ yielding a {\em (total)
spacetime connection\/} $K$, with coordinate expression
\beq
K_i{}^h{}_j = K\Nat{}_i{}^h{}_j \,,
\quad
K_j{}^h{}_0 = K_0{}^h{}_j = K\Nat{}_0{}^h{}_j + \tfrac12 \, F^h{}_j
\,,
\quad 
K_0{}^h{}_0 = K\Nat{}_0{}^h{}_0 + \tfrac12 \, F^h{}_0 \,.
\eeq
This turns out to be a time preserving torsion free linear connection
of the tangent space of spacetime, which still fulfills the properties
that we have assumed for $K\Nat$.

The fibring of spacetime, the total spacetime connection and
the spacelike metric yield, in a covariant way, a 2--form
$\Ome : J_1\bE \to \Lam^2 T^*J_1\bE$ of phase space, with coordinate
expression
\beq
\Ome = G^0_{ij} \,
\big(dx^i_0 - (K_\lam{}^i{}_0 + K_\lam{}^i{}_h \, x^h_0)\,
dx^\lam\big) \wed (dx^j - x^j_0 \, dx^0) \,.
\eeq
This is a {\em cosymplectic form\/}
\cite{CandeLLac92,deLSar93,deLTuy96,Lib59,LibMar87}, i.e. it fulfills
the following properties: 1) $d \Ome = 0$, 2) 
$dt \wed \Ome^n : J_1\bE \to \BT \ten \Lam^nT^*J_1\bE$ is a scaled
volume form of $J_1\bE$. Conversely, the
cosymplectic form
$\Ome$ characterises the spacelike metric and the total spacetime
connection. Moreover, the closure of $\Ome$ is equivalent to the
conditions that we have assumed on $K$.

There is a unique second order connection \cite{ManMod83}
$\gam : J_1\bE \to \BT^* \ten TJ_1\bE$, such that $i_\gam \Ome = 0$.
We assume the generalised {\em Newton's equation\/}
$\nab[\gam] j_1s = 0$ as the equation of motion for classical
dynamics \cite{JadMod94b,MasPag91,MasPag93,VerFad81}.

We can also obtain this equation by a Lagrangian formalism according
to a cohomological procedure in the following way
\cite{Kru90,JadJanMod98,ModVit96}. The cosymplectic form $\Ome$
admits locally potentials of the type
$\The : J_1\bE \to T^*\bE$, defined up to a closed form of the type
$\alp : \bE \to T^*\bE$, which are called
{\em Poincar\'e--Cartan forms\/} \cite{Gar74,GolSte73,ModVit96}.
Each Poincar\'e--Cartan form $\The$
splits, according to the splitting of $T^*\bE$ induced by $J_1\bE$,
into the horizontal component
$\cL : J_1\bE \to T^*\bT$, called {\em Lagrangian\/}, and the
vertical component
$\cP : J_1\bE \to V^*\bE$, called {\em momentum\/}.
These components are observer independent, but depend on the chosen
gauge of the starting Poincar\'e--Cartan form. On the other hand,
given an observer $o$, each Poincar\'e--Cartan form $\The$
splits, according to the splitting of $T^*\bE$ induced by $o$,
into the horizontal component
$- \cH[o] : J_1\bE \to T^*\bT$, called observed {\em Hamiltonian\/},
and the vertical component
$\cP[o] : J_1\bE \to V^*\bE$, called observed {\em momentum\/}.
Moreover, the horizontal component of $\Ome$, according to the
splitting of $T^*J_1\bE$ induced by $J_2\bE$, is the map
$\cE = G\Fla (\nab[\gam]) : J_2\bE \to \BT^* \ten V^*\bE$,
which turns out to be the Euler--Lagrange operator associated with
$\cL$. We have the coordinate expressions
\beq
\cL = (\tfrac12 \, G^0_{ij} \, x^i_0 x^j_0 + A_i \, x^i_0 + A_0)
\, dx^0, \quad
\cP = (G^0_{ij} \, x^j_0 + A_i) \, (dx^i - x^i_0 \, dx^0) \,,
\eeq 
and, in a chart adapted to $o$,
\beq
\cH[o] = (\tfrac12 \, G^0_{ij} \, x^i_0 x^j_0 - A_0) \, dx^0,
\quad
\cP[o] = (G^0_{ij} \, x^j_0 + A_i) \, dx^i \,,
\eeq
where
$A \equiv o^* \The$.

The cosymplectic form $\Ome$ yields in a covariant way the
Hamiltonian lift of functions $f: J_1\bE \to \R$ to vertical vector
fields $H[f] : J_1\bE \to VJ_1\bE$; consequently, we obtain the
Poisson bracket $\{f,g\}$ between functions of phase space. Given an
observer, the law of motion can be expressed, in a non covariant way,
in terms of the Poisson bracket and the Hamiltonian.

More generally, chosen a time scale
$\tau : J_1\bE \to T\bT$, the cosymplectic form
$\Ome$ yields, in a covariant way, the Hamiltonian lift of functions
$f$ of phase space to vector fields
$H_\tau[f] : J_1\bE \to TJ_1\bE$, whose time component is $\tau$. In
view of our developments in the Quantum Theory, we prove that 
$H_\tau[f]$ is projectable on a vector field
$X[f] : \bE \to T\bE$ if and only if the following conditions
hold: i) the function $f$ is quadratic with respect to the affine
fibres of $J_1\bE \to \bE$ with second fibre derivative $f'' \ten G$,
where $f'' : \bE \to T\bT$, ii) $\tau = f''$.
A function of this type is called {\em special quadratic\/} and has
coordinate expression of the type
\beq
f = \tfrac12 f^0 \, G^0_{ij} x^i_0 x^j_0 + f^0_i \, x^i_0 + \obf \,,
\qquad
{\rm with}
\qquad
f^0, f^0_i, \obf : \bE \to \R \,.
\eeq

The vector space of special quadratic functions is not closed under
the Poisson bracket, but it turns out to be an $\R$--Lie algebra
through the covariant {\em special bracket\/}
\beq
\lsqu f, g\rsqu = \{f,g\} + \gam(f'').g - \gam(g'').f \,.
\eeq
We have the subalgebra of {\em quantisable\/} functions whose time
component factorises through $\bT$, the subalgebra of functions
whose time component is constant, the subalgebra of {\em affine\/}
functions whose time component vanishes and the abelian subalgebra of
{\em spacetime\/} functions which factorise through $\bE$. In
particular, the Hamiltonian is a quantisable function, the components
of the momentum are affine functions and the spacetime coordinates
are spacetime functions.

Moreover, the map
$f \mto X[f]$ turns out to be a morphism of Lie algebras. The
coordinate expression of the tangent lift of $f$ is
$X[f] = f^0 \, \der_0 - f^i \, \der_i$.
\subsection{Quantum theory}
\label{Quantum theory}

The quantum framework is described in the following way.

A {\em quantum bundle\/} is defined to be a 1--dimensional complex
vector bundle over spacetime $\bQ \to \bE$ equipped with a Hermitian
metric
$h : \bQ \ucar{\bE} \bQ \to \BC \ten \Lam^n V^*\bE$ with values in the
complexified volume forms of the fibres of spacetime. We shall refer
to normalised local bases $b$ of $\bQ$ and to the associated complex
coordinates $z$; accordingly, the coordinate expression of a {\em
quantum section\/} is of the type $\Psi = \psi \, b$, with
$\psi : \bE \to \BC$.

We consider also the {\em extended quantum bundle\/}
$\bQ\Upa \to J_1\bE$, by taking the pullback of $\bQ \to \bE$, with
respect to the map $J_1\bE \to \bE$. A system of connections of $\bQ$
parametrised by the sections of $J_1\bE \to \bE$ induces, in a
covariant way, a connection of $\bQ\Upa$, which is called {\em
universal\/} \cite{Gar72,ManMod83,JadJanMod98}. A characteristic
property of the universal connection is that its contraction with any
vertical vector field of the bundle
$J_1\bE \to \bE$ vanishes.

A {\em quantum connection\/} is defined to be a connection $\cyq$ of
the extended quantum bundle, which is Hermitian, universal and whose
curvature is $R[\cyq] = \rmi \, \Ome$. We stress that $\tfrac{1}{\h}$
has been incorporated in $\Ome$ through the re--scaled metric
$G$. 
In a chart adapted to the observer $o$, the coordinate expression of a
quantum connection is locally of the type
\beq
\cyq_0 = - \cH[o] \,,
\qquad
\cyq_i = \cP[o] \,,
\qquad
\cyq^0_i = 0 \,,
\eeq
where the choice of the potential $A[o]$ is locally determined by
$\cyq$ and by the quantum base $b$. A quantum connection exists if
and only if the cohomology class of $\Ome$ is integer; the
equivalence classes of quantum bundles equipped with a quantum
connection are classified by the cohomology group $H^1\big(\bE,
U(1)\big)$ \cite{Vit99b}.

We stress the minimality of our quantum bundle and quantum connection.

Let us assume a quantum bundle equipped with a quantum connection.

Any other quantum object is obtained, in a covariant way, from this
quantum structure. The quantum connection lives on the extended
quantum bundle, while we are looking for further quantum objects
living on the original quantum bundle. This goal is successfully
achieved by a {\em method of projectability\/}: namely, we look for
objects of the extended quantum bundle which are projectable to the
quantum bundle and then we take their projections. Indeed, our method
of projectability turns out to be our way of implementing the
covariance of the theory; in fact, it allows us to get rid of the
family of all observers, which is encoded in the quantum connection
(through $J_1\bE$).

J. Jany\v{s}ka \cite{Jan98,Jan00} has proved that all covariant
quantum Lagrangians of the quantum bundle are proportional to
\beq
\EL[\Psi] = dt \wed 
\left(h(\Psi, \rmi \banab \Psi) + h(\rmi \banab \Psi, \Psi)
- (\baG \ten h)(\venab \Psi, \venab \Psi)
+ k \, r \, h(\Psi, \Psi) \right) \,,
\eeq
where $k$ is an arbitrary real factor, $\banab$ denotes the
covariant differential with respect to time induced by the phase
space, $\venab$ denotes the vertical covariant differential and
$r : \bE \to \R \ten \BT^*$ is the scalar curvature of the spacelike
metric $G$. Thus, $k$ remains undetermined in our scheme. Several
authors have tried to determine this factor via Feynmann's path
integral approach, but they found different results, according to
different ways to perform the integral
\cite{Che72,DeW57,Fey48,KapMaiHel98}.

The standard Lagrangian formalism yields, from the above covariant
quantum Lagrangian, the covariant quantum $(n+1)$--momentum, the
covariant Euler--Lagrange equation and the covariant conserved
probability current. These objects can also be obtained directly in
terms of covariant differentials through the quantum connection, by
means of the projectability method. The coordinate expression of the
Euler--Lagrange equation is
\beq
\ES. \psi \equiv \big(\obnab_0 + \tfrac12 \, \sqrtgo\big) \psi - \rmi
\tfrac12 \big(\obDel_0 - k \, r_0\big) \, \psi = 0 \,,
\eeq
where
\beq
\obDel_0 \equiv G^{hk}_0 \, \obnab_h \, \obnab_k + K_h{}^k{}^h \,
\obnab_k
\,,
\qquad
\obnab_\lam \equiv \der_\lam - \rmi A_\lam \,,
\eeq
denote the Laplacian and the covariant differential induced by the
connections $\cyq, K$ and by the observer attached to the spacetime
chart. J. Jany\v{s}ka \cite{Jan00} has proved that any covariant
Schr\"odinger equation is of the above type, hence it is the
Euler--Lagrange equation associated with a covariant Lagrangian. We
assume the above equation as the {\em quantum dynamical equation\/}.

Next, we classify the Hermitian vector fields of the extended
quantum bundle, which are projectable to the quantum bundle. We find
that the projected vector fields
$Y : \bQ \to T\bQ$ of the quantum bundle, called
{\em quantum vector fields\/}, constitute a Lie algebra naturally
isomorphic to the Lie algebra of quantisable functions. The
coordinate expression of the quantum vector field associated with the
quantisable function $f$ is
\beq
Y[f] = f^0 \, \der_0 - f^j \, \der_j 
+ \left(\rmi (f^0 \, A_0 -
f^h \, A_h + \obf) - \tfrac12 \, \Div X[f]\right) \, z \, \der z \,,
\eeq
where
$\Div X[f] =
f^0  \, \frac{\der_0\sqrtg}{\sqrtg} - \frac{\der_j(f^j\sqrtg)}{\sqrtg}
$. 

The quantum vector field $Y[f]$ acts on the sections $\Psi$ of the
quantum bundle via the associated Lie derivative 
$Z[f] \byd \rmi \, Y[f]_\bullet \, $.
In particular, we obtain
\beq
Z[\cH_0](\Psi) =
\rmi \, (\der_0 + \tfrac12 \frac{\der_0\sqrtg}{\sqrtg}) \, \psi \,
b \,,
\quad
Z[\cP_j](\Psi) =
- \, \rmi \, (\der_j + \tfrac12 \frac{\der_j\sqrtg}{\sqrtg}) \, \psi
\, b \,.
\eeq

Next, we consider the pre--Hilbert {\em functional quantum bundle\/}
$\bH \to \bT$ over time, whose infinite dimensional fibres are
constituted by the sections of the quantum bundle at a given time and
with compact support. The quantum dynamical operator $\ES$ can be
regarded as a covariant differential $\nab[\chi]$ of the functional
quantum bundle; hence, the quantum Lagrangian yields a lift of the
quantum connection $\cyq$ of the extended quantum bundle to a
connection $\chi$ of the functional quantum bundle.

Moreover, we can see that, if $f$ is a quantisable function, then
\beq
\hat{f} = \rmi (Y[f]_\bullet - f'' \con \nab[\chi])
\eeq
is the unique combination of $Z[f]$ and $\nab[\chi]$, which yields an
operator acting on the fibres of the functional quantum bundle. We
have the following coordinate expression
\beq
\hat{f}(\Psi) 
= \big(- \tfrac12 \, f^0 \,  \obDel{}_0
- \rmi \, f^j \, \obnab_j
+ \obf + \tfrac12 \, k \, f^0 \, r_0 
- \rmi \, \tfrac12 \, \frac{\der_j (f^j \, \sqrtg)}{\sqrtg} \big) \,
\psi \, b \,.
\eeq
The map
$f \mto \hat{f}$ is injective. Moreover, $\hat{f}$ is Hermitian. 

We assume $\hat{f}$ to be the Hermitian {\em quantum operator\/}
associated with the quantisable function $f$. This is our {\em
correspondence principle\/}.

We define the commutator of Hermitian fibred operators $h, k$ of the
functional quantum bundle by 
$[h \,, \, k] \byd - \rmi \, (h k - kh)$.
Then, for each quantisable
functions $f, g$, we obtain the formula
\beq
[\hat{f} \,,\, \hat{g}] = \widehat{[f \,,\, g]} 
+ \big[(g'' \ten Y[f]_\bullet - f'' \ten Y[g]_\bullet) \,,\, \ES\big]
\,.
\eeq
The second term in the above formula is the obstruction for the map 
$\hat{} : f \mto \hat{f}$ to be a morphism of Lie algebras.
There is any substantial physical reason by which the map
$\hat{}$ should be a morphism of Lie algebras? On the other hand, the
restriction of the map $\hat{}$ to the subalgebra of affine
functions yields an injective morphism of Lie algebras.

The Feynmann path integral formulation of Quantum Mechanics
\cite{Che72,Fey48} can be naturally expressed in our formalism; in
particular, the Feynmann amplitudes arise naturally via parallel
transport with respect to the quantum connection \cite{JadMod94b}. So
the Feynmann path integral can be regarded as a further way to lift
the quantum connection $\cyq$ to a functional quantum connection.

In the particular case when spacetime is flat, our quantum
dynamical equations turns out to be the standard Schr\"odinger
equation and our quantum operators associated with spacetime
coordinates, momenta and energy coincide with the standard operators.

Therefore, all usual examples of standard Quantum Mechanics are
automatically recovered in our covariant scheme.

\myskip

The above procedure can be easily extended to classical and quantum
multi--body systems.

The above covariant theory can be extended to particles with spin; in
this way, we obtain a generalised Pauli equation and all that.

Several techniques of the above theory (including the Lie algebra of
quantisable functions and the corresponding Lie algebra of quantum
vector fields) can be reproduced on a Lorentz manifold in a covariant
way in the sense of Einstein. However, we do not know so far how to
achieve a Hilbert stuff in this contest. It is possible that this
problem has no solution (out of the Quantum Field Theory), as it is
commonly believed.
\section{Comparison}
\label{Comparison}

Covariant Quantum Mechanics (CQM) has several points in common with
Geometric Quantisation (GQ) \cite{AbrMar78,Gar79,Sni80,Woo92}. The
differences between the two theories arise from the fact that their
basic goals are different: quantisation procedure and covariant
formulation, respectively.
\subsection{Quantisation}
\label{Quantisation}

GQ can be regarded as a general programme aimed at ``quantising" a
classical system. More  precisely, GQ is aimed at establishing a
procedure in order to represent an algebra of functions of a classic
symplectic manifold into a Hilbert space, according to some
reasonable rules.

Perhaps, the original notion of ``canonical quantisation" goes back to
P. M. Dirac \cite{Dir67}. The first rigorous mathematical formulation
of the notion of ``geometric pre-quantisation" was due to I. E. Segal
\cite{Segal60}; later J. M. Soriau \cite{Sou70} and B. Kostant
\cite{Kos70} founded the Geometric Quantisation. This theory
has been refined by several authors, see \cite{Bla74,Bla77,LanLin91}
and
\cite{Got86}. For instance, see \cite{AbrMar78}, a ``full
quantisation" of a symplectic manifold $(M,\ome)$ is defined to be
a pair $(\cH, \del)$ where $\cH$ is a separable
complex Hilbert space and $\delta$ is a map taking functions $f\in
\Cin(M)$ to self adjoint operators $\del_f$ of $\cH$ such that
\begin{enumerate}
\item $\delta$ is $\R$-linear,

\item $\del_1  = \id_\cH$,

\item $[\del_f, \del_g] = \rmi \, \h \, \del_{\{f,g\}}$,

\item if $\cA \in \Cin(M)$ is a complete subalgebra,
i.e. if its centraliser with respect to the Poisson bracket is $\R$,
then $\del_\cA$ acts irreducibly on $\cH$.

\end{enumerate}

There are {\em no go\/} theorems stating that there are no such
quantisations in several situations
\cite{AbrMar78,Gar79,Got86,Got95,Got99,GotGraGru99,GotGruTuy96,
GotGru97,GotGru99,GotGruHur96,Sni80,Woo92};
among them we mention the famous Groenewold - Van Hove theorem for the
symplectic manifold
$(\R^{2n},\ome)$. If there is no full quantisation, then one looks for
a subalgebra $\cO \in \Cin(M)$ to be quantised.

Since the requirement of a quantisation is too restrictive, one
defines, as first step, a {\em pre-quantisation\/} by requiring
just properties $(1,2,3)$. A pre-quantisation, called the {\em
Dirac problem\/}, exists for every symplectic manifold, whose
symplectic form defines an integer cohomology class.

\myskip

In some respects, the aim of CQM is not the quantisation of a
classical system. More precisely, we are just looking for a covariant
formulation of the standard Quantum Theory \cite{Dir67,Mes61}. On the
other hand, any quantum measurement is eventually constituted by
classical observations, so we need to consider a classical spacetime
as background of the Quantum Theory. Then, this background structure
plays an important role in the Quantum Theory. But our heuristic
geometric techniques are partially different from those of
representations of Lie algebras. We observe also that in our
formulation the classical spacetime and its structures, rather than
the classical dynamics, determine the Quantum Theory.

Eventually, we do obtain a correspondence principle, which is a
consequence of a classification theorem and not a postulate. But, this
can be regarded as a quantisation only partially.
\subsection{Covariance}
\label{Covariance}

The standard literature on GQ is not concerned with the special or
general relativistic covariance of the theory. Indeed, the language
of GQ is geometric, hence coordinate free; but this is not sufficient
for attaining the covariance. In fact, in the standard literature on
GQ, a given frame of reference is implicitly assumed.

\myskip

On the other hand, CQM looks for a formulation of standard Quantum
Mechanics, which be manifestly covariant (with respect to all frames
of references, including accelerated frames), in the spirit of
General Relativity.

Indeed, it would be natural to take a curved Lorentz manifold as
classical background spacetime for such a theory. However, it is well
known that there are serious physical difficulties to formulate the
Special or General Relativistic Quantum Mechanics in this framework;
actually, these difficulties led to the Quantum Field Theory. On the
other hand, we realised that it is possible to keep the framework of
Quantum Mechanics (Schr\"odinger and Pauli equations and all that) and
formulate our covariant theory in a curved spacetime with absolute
time and spacelike Riemannian metric. This approach stands in between
the standard non relativistic Quantum Mechanics and a possible general
relativistic Quantum Mechanics. In fact, our classical spacetime
supports accelerated frames and several features of General
Relativity (including the geometric interpretation of the
gravitational field), but misses all features strictly related to the
Lorentz metric (including the finite speed of signals).

In the flat case, this setting allows us to recover the standard non
relativistic Quantum Mechanics. In the curved case, it suggests
several new interpretations and techniques which might be possibly
useful for a ``true" general relativistic theory.

Thus, the covariance is the leading principle of our theory. Indeed,
it turns out to be a powerful heuristic guide, as in all general
relativistic theories. All main differences between our theory and GQ
are related to the covariance.

Our covariant approach can be compared with a large literature
dealing with Galileian General Relativity 
\cite{DomHor64,Duv93,DuvBurKun85,DuvGibHor91,DuvKun84,Ehl89,Hav64,
HorPir73,Kun72,Kun74,Kun76,Kun84,LebLev73,Lev71,Man79,Tra63,Tra66}
and covariant formulations of Quantum Mechanics 
\cite{Duv85,Fan93,Fan94,Hor92,HorRot81,Kuc80,Kyp87,Pau58,SchPle77,
Tul85}.
\subsection{Generality}
\label{Generality}

The starting programme of GQ is quite general and is based on weak
assumptions. In fact, GQ deals just with a symplectic manifold without
further structure. On the other hand, strong symmetries of the
framework are usually assumed and specified case by case.

\myskip

CQM starts with a fibred manifold equipped with a spacelike metric and
a fibre preserving linear connection, which fulfill a natural
condition. In our opinion, the above geometric structure well reflects
the physical features occurring in all examples of interest for
Quantum Mechanics and yields in a functorial way any further object
which is needed for the development of the classical theory
(including the cosymplectic structure).

Thus, the CQM deals with a type of model more specific than that of
GQ. On the other hand, the large generality of GQ is a beautiful
mathematical feature, which, in practice, cannot be physically
implemented in full extent. Actually, perhaps all concrete examples
of physical interest that can be treated in the framework of GQ can
be regarded as particular cases of our model.
\subsection{Role of time}
\label{Role of time}

In GQ, time is essentially an exterior parameter. This theory
basically deals with classical and quantum systems which do not
depend explicitly on time. So, the starting classical configuration
space is a manifold $\bS$ which does not ``include" time. If the
theory needs to consider time, then it refers to the product manifold
$\bE \equiv \bT \car \bS$; the fact that spacetime is a product
manifold means that a global observer has been implicitly chosen.

\myskip

In CQM, the requirement that the theory be observer independent
imposes that spacetime ``includes" time but be not naturally split
into space and time.

In Einsteinian General Relativity, spacetime $\bE$ yields no
observer independent time $\bT$ and space $\bS$, hence we have no
observer independent projections $\bE \to \bT$ and $\bE \to \bS$. In
our Galileian General Relativity, spacetime $\bE$ is equipped
with an observer independent time $\bT$ and projection $\bE \to \bT$,
but we have no observer independent space $\bS$ and projection $\bE
\to \bS$. In non relativistic GQ, spacetime $\bE$ is equipped with
time $\bT$, space $\bS$, and projections 
$\bE \to \bT$ and $\bE \to \bS$.

The further developments of our theory respect the starting
assumption on the existence of absolute time without a preferred
splitting of spacetime. Thus, all peculiar features of our theory
follow from the covariance through the role of time. In particular,
the cosymplectic structure of our phase space, the
universality of the quantum connection, the method of projectability,
the absence of the problem of polarisations and the construction of
classical and quantum Hamiltonians are related to the role of time.
\subsection{Phase space}
\label{Phase space}

In GQ the phase space is, in principle, any manifold supporting a
symplectic form. Usually, the cotangent manifold of a manifold plays
the role of phase space in virtue of the fact that it carries a
canonical symplectic form. Thus, phase space has even dimension.

\myskip

In CQM phase space is constituted by the first jet of sections of
the spacetime fibred over time \cite{KolMicSlo93,ManMod83,Sau89}. This
choice is essential for the covariance of the theory. Thus, the phase
space has odd dimension. Actually, the techniques related to even and
odd phase spaces, respectively, present important differences.

On the other hand, any observer induces an affine fibred isomorphism
of the first jet bundle with the vertical tangent bundle of spacetime
(up to a time--scale factor); moreover, the spacelike metric induces a
linear fibred isomorphism of the vertical tangent bundle with the
vertical cotangent bundle of spacetime. Thus, breaking the
covariance, the choice of an observer and the reference to the
spacelike metric allow us to compare our phase space with that of GQ.
\subsection{Symplectic and cosymplectic structures}
\label{Symplectic and cosymplectic structures}

In GQ the basic geometric structure of Classical Mechanics is
constituted by a symplectic form and a Hamiltonian function of phase
space. In principle, nothing else is necessary; in practice, one adds
the fibring of the phase space over the configuration space and a
suitable group of symmetry.

\myskip

In CQM the geometric structure of Classical Mechanics is constituted
by the spacetime fibred over time, the spacelike metric and the
spacetime connection. These objects yield a cosymplectic form in a
covariant way.

Any observer yields, by pullback and vertical restriction, a
Riemannian symplectic form of the fibres of the vertical tangent
bundle of spacetime. In this way we recover the analogue of the
symplectic form of GQ. However, we stress that this symplectic form is
not covariant and carries less information than the original
cosymplectic form.

Furthermore, the cosymplectic form and the choice of an observer
yield a classical Hamiltonian function.
Thus, in CQM the cosymplectic form encodes the Hamiltonian (but an
observer is needed to extract it).
\subsection{Classical Lie brackets}
\label{Classical Lie brackets}

In the original programme of GQ the classical Lie algebra to be
quantised is the Poisson Lie algebra of all functions of the phase
space, whose bracket is associated with the symplectic form.
Actually, we have mentioned before that some obstructions to this
quantisation programme occur, but there is no subalgebra $\cO$ that
can be consistently considered for all cases.

\myskip

In CQM we do define the Poisson Lie algebra of all functions of
phase space, whose bracket is associated with the cosymplectic form.
However, we are only partially interested in this algebra. Indeed, we
exhibit a new Lie algebra of special quadratic functions, which is
involved in our Quantum Theory. This Lie algebra includes all
functions which are usually quantised in the standard approaches.
\subsection{Symmetries}
\label{Symmetries}

In GQ the conserved quantities associated with the group of symmetries
of the classical system are not necessarily quantisable. For a system
whose phase space is a co-adjoint orbit for some group, one possibly
imposes that the generators of the group are quantisable and act
irreducibly on the Hilbert space. This is done in order to establish
a correspondence between ``elementary systems'' at the classical and
quantum levels, and can be considered as a sort of irreducibility
condition.

\myskip

In CQM there is no need for any specific group of symmetries acting on
the classical system. Actually, the procedure of quantisation does
not depend on such a group. On the other hand, the possible classical
symmetries yield interesting consequences, including the momentum map 
\cite{AbrMar78,MarRat95,SalVit99}. In particular, all symmetries of
the classical structure yield conserved quantisable functions
\cite{SalVit99}.
\subsection{Quantum structure}
\label{Quantum structure}

In GQ the quantum bundle is assumed to be a Hermitian line bundle over
phase space. Moreover, the quantum bundle is assumed to be
equipped with a Hermitian connection whose curvature is proportional
to the classical symplectic form.

\myskip

In CQM the quantum bundle is assumed to be a Hermitian line bundle
over spacetime. Moreover, the extended quantum bundle is assumed to be
equipped with a Hermitian connection whose curvature is universal and
proportional to the classical cosymplectic form.

Thus the novelties of CQM consist in the following minimal
assumptions: the quantum bundle lives on spacetime and not on the
phase space, the quantum connection is universal.
\subsection{Polarisation and projection method}
\label{Polarisation and projection method}

In GQ one realises that the base space of the quantum bundle is too
big in order to obtain an irreducible representation of the classical
Poisson Lie algebra and to fulfill the uncertainty principle. Then,
one looks for a polarisation $P$, that is for a Lagrangian subbundle
$P$ of the complexified tangent bundle of the phase space $M$, such
that $D_\BC = P \cap \baP$ has constant rank and $P$, $P + \baP$ are
closed under the Lie bracket; moreover, the polarisation is said to be
reducible if the quotient of the phase space $M$ by the distribution
$D$ exists and the canonical projection $\pi : M \to M/D$ is a
submersion. Once a polarisation is chosen, we can consider the
polarised sections of the quantum bundle, that is the sections whose
covariant derivative with respect to every vector field of the
polarisation vanish. The polarised sections should yield the Hilbert
space with the correct size. Actually, the problem of finding
polarisations is very hard in practice, should be faced case by
case and leads to a lot of complications and ambiguities, where the
beauty of the original programme misses over considerably.

\myskip

In CQM we have, in a covariant way, an implicit natural polarisation,
namely the vertical polarisation.

The quantum connection is the only source of all further quantum
objects, such as quantum dynamical equation, probability current,
quantum operators and so on.

On the other hand, the quantum connection lives on the extended
quantum bundle, while we are looking for further quantum objects
living on the original quantum bundle. This goal is successfully
achieved by the method of projectability.

Thus, in simple words, the difficult search for the inclusion of a
polarisation (which should be performed case by case) is substituted
by the easy search for projectable objects (which is successful and
can be performed in general).
\subsection{Half--densities and half--forms }
\label{Half--densities and half--forms}

In GQ, once a polarisation $P$ is chosen, we take the polarised
sections of the quantum bundle. The problem is how to build a Hilbert
space. For instance, if $P$ is reducible, then the Hermitian product
$h(s_1, s_2)$ of two polarised sections can be understood as
a function on the quotient space $M/D$; but a problem arises because
$M/D$ has no natural volume element, which is necessary to define the
Hilbert space of $L^2$--polarised sections. One way to remedy this
problem is to tensor out the sections by the half--densities
associated to $D$ and to use the natural partial flat connection of
$D$ to build the Hilbert space. Unfortunately, even for the simplest
physical systems, the results do not agree with those found in
Quantum Mechanics.

Therefore, a further modification of the theory is needed: here is
where half-forms come into play. They are defined through a
metaplectic structure for the bundle of frames of the polarisation
$P$; this imposes new conditions, since metaplectic structures do not
exist in general. Even further problems arise, since it may happen
that the tensor product of the quantum bundle with the bundle of
half--forms has no polarised section at all.

\myskip

In CQM we do not see any trouble concerning half--densities and all
that. It seems that this convenient feature depends on the fact that
CQM gets rid of all problems concerning polarisations.

In the first version of CQM the theory assumed a $\BC$--valued
Hermitian metric of the quantum bundle. In this case the theory, in
order to prove that quantum operators are Hermitian, needed to use
half--densities.

In the recent version of CQM the theory assumes a 
$(\BC \ten \Lam^nV^*\bE)$--valued Hermitian metric of the quantum
bundle. In this case the theory uses just sections of the quantum
bundle and does not need half--densities at all.
\subsection{Schroedinger equation}
\label{Schroedinger equation}

In GQ there is no clear Schr\"odinger equation in an explicit
formulation ready to be directly compared with the Schr\"odinger
equation of standard Quantum Mechanics. The Schr\"odinger equation in
this framework is understood as the infinitesimal generator of the
flow of the Hamiltonian acting on the Hilbert space. The problem is
that in general the Hamiltonian does not preserve the chosen
polarisation and therefore we are led to compare different
polarisations, this is handled through the Blattner-Kostant-
Sternberg kernel machinery, see \cite{Bla77}, which is extremely
cumbersome. Even here one encounters new problems since there are
obstructions discovered by Blattner \cite{Bla77} in order to
construct the BKS kernel.

\myskip

In CQM we have an explicit, covariant and intrinsic Schr\"odinger
equation, which is immediately comparable with the standard
Schr\"odinger equation \cite{Dir67,Mes61} (see also 
\cite{Doe98,DoeGol92,DoeGol94,DoeGol96,DoeGolNat99,DoeHen97,
DoeNat96}).
Moreover, we prove that it comes from a quantum Lagrangian.
\subsection{Hilbert space and Hilbert bundle}
\label{Hilbert space and Hilbert bundle}

GQ and standard Quantum Mechanics deal with a Hilbert space, which
usually consists of $L^2$ sections of the quantum bundle.

\myskip

CQM deals with a Hilbert bundle over time. This formulation is
unusual but does not seem to be really in contrast with standard
Quantum Mechanics.

Once more, this novelty is related to the explicit role of time in the
CQM.
\subsection{Feynmann path integral}
\label{Feynmann path integral}

In CQM the Feynmann amplitudes appear very naturally in terms of
parallel transport with respect to the quantum connection. In fact,
the classical Lagrangian turns out to play the role of local symbol
of the quantum connection of the extended quantum bundle over time.

The proof of the equivalence of the Feynmann path integral with the
covariant Schr\"odinger equation has not yet been worked out in
detail.
\subsection{Energy}
\label{Energy}

Most of the practical difficulties of GQ run around the quantisation
of energy. Actually, the classical Hamiltonian function has to be
explicitly postulated and is not encoded in the basic structure of
spacetime. Moreover, the energy requires a treatment quite different
from the simpler approach required by other observables such as
spacetime coordinates and momentum.

\myskip

Even in CQM the energy has a special role, but no hard problems arise
in this respect. First of all, we stress that energy is encoded in
the basic geometric structure of spacetime and that it appears
explicitly, in a non covariant way, by means of the choice of an
observer.

If one accepts the point of view of Covariant Quantum Mechanics, maybe
one can understand the difficulties of GQ from this perspective. In
fact, in practice what is done in GQ is to take the vertical tangent
(or cotangent) space of spacetime as phase space, instead of the
first jet space; accordingly, GQ tries to formulate and quantise
energy by methods related to vertical subspace. On the other hand, in
CQM, it is very clear that energy is related to the horizontal aspect
of phase space. We could roughly say that the vertical aspect of
phase space is essentially related to the static geometry of
spacetime, while the horizontal aspect is related to the dynamics.
This observation can also be analysed in terms of the Lie algebra of
quantisable functions and their tangent lift; actually, the
quantisable functions dealing with the vertical aspects of spacetime
constitute the subalgebra of affine functions, while energy is a
quadratic function. In CQM, all quantisable functions can quantised
on the same footing.
\subsection{Examples}
\label{Examples}

In GQ it is not granted that every reasonable physical example can be
worked out. Actually, there are few examples that have been
successfully solved.

CQM reduces to standard Quantum Mechanics in the flat case. Hence, in
CQM, all standard physical examples can be formulated; moreover, the
Schr\"od\-inger equation and quantum operators corresponding to the
quantisable functions can be explicitly and immediately computed. Of
course, the integration of the Schr\"odinger equation and the
computation of the energy spectrum is an analytical question which
should be faced case by case.


\end{document}